\documentclass[%
 reprint,
 amsmath,amssymb,
 aps,
 prb,
]{revtex4-2}

\usepackage{graphicx}% Include figure files
\usepackage{dcolumn}% Align table columns on decimal point
\usepackage{bm}% bold math
\usepackage{hyperref}
\usepackage{amsthm}

\usepackage{subcaption}
\usepackage{ragged2e}
\usepackage{mathdots}

\begin{document}

\captionsetup{justification=justified, singlelinecheck=false}

\preprint{APS/123-QED}

\title{Quantum Cosmology on Quantum Computer}% Force line breaks with \\
%	\thanks{A footnote to the article title}%

\author{Chih-Chien Erich Wang}
%\email{F05244004@ntu.edu.tw}
 %\altaffiliation[Also at ]{Physics Department, National Taiwan University.}%Lines break automatically or can be forced with \\
\author{Jiun-Huei Proty Wu}%
 \email{jhpw@phys.ntu.edu.tw}
\affiliation{%
	Department of Physics, Institute of Astrophysics, and National Center for Theoretical Sciences, National Taiwan University, Taipei 10617, Taiwan
}%

\date{\today}% It is always \today, today,

\begin{abstract}
	With physical quantum computers becoming increasingly accessible, research on their applications across various fields has advanced rapidly. In this paper, we present the first study of quantum cosmology conducted on physical quantum computers, employing a newly proposed Hybrid Quantum-Classical (HQC) algorithm rather than the commonly used Variational Quantum Eigensolver (VQE). Specifically, we solve a constrained Hamiltonian equation derived by quantizing the Friedmann equation in cosmology. 
	To solve this constraint equation $H|\psi \rangle = 0 $, where $H$ is a Hamiltonian operator and $|\psi \rangle \equiv |\psi(\theta) \rangle$ is the wave function of phase angle $\theta$ describing the cosmic quantum state, 
	we iteratively use the quantum computer to compute the eigenvalues of $\langle \psi |H |\psi \rangle$, while a classical computer manages the underlying probability density function within the Probabilistic Bisection Algorithm (PBA) to update $\theta$ until the solution of $\langle \psi |H |\psi \rangle = 0$ is achieved to a desired accuracy. Executing our algorithm on IBM’s quantum computers, we attain a high-precision solution for $ \theta$, achieving approximately $1\%$ error.
\end{abstract}

\maketitle

%\tableofcontents

%\section{\label{sec:level1}First-level heading:\protect\\ The line
%break was forced \lowercase{via} \textbackslash\textbackslash}
\section{Introduction}

Since Feynman first introduced the concept of quantum computing in the 1980s \cite{Feynman:85}, researchers have sought to identify problems particularly suited to quantum computation. In the 1990s, several problems were identified for which quantum algorithms offer exponential speedups over their classical counterparts. Notable examples include the Deutsch–Jozsa algorithm \cite{1992RSPSA.439..553D}, Quantum Phase Estimation \cite{kitaev1995quantum}, polynomial-time algorithms for prime factorization and discrete logarithms \cite{Shor_1997}, and database search \cite{grover1996fast}.

In the current era of Noisy Intermediate-Scaled Quantum (NISQ) Device, the noisy-ness problem of the physical qubits makes it challenging to identify useful applications of quantum computers on physics problems. To mitigate such limitation, the Hybrid Quantum-Classical (HQC) algorithms, which combine the strengths of both classical and quantum computers, are proved to be useful. One of the most well known and commonly used HQC algorithms is the Variational Quantum Eigensolver (VQE) \cite{osti_1623945,McClean_2016,Romero_2018,Google2020,IBM2017,PhysRevX.6.031007,ISI:000524530000001,doi:10.1021/acs.jpca.0c09530}.

In this paper, we solved a constrained equation in Quantum Cosmology by employing a new type of HQC algorithm. We used quantum computers to solve for the eigenvalue of a system, and the classical computers to estimate and update the related parameters.

We organize this paper as follows.
In Section~\ref{Sec.QuantumCosmology}, we first introduced fundamental equations in Quantum Cosmology and derived the constraint equation, which would be solved by a combination of quantum computers and classical computers. 
In Section~\ref{Sec.Methodology}, we introduced our framework through which to solve the equation.
In Section~\ref{Sec.ResultandDiscussion}, we presented our results with discussions.
Finally we concluded our work in Section~\ref{Sec.Conclusion}.

\section{\label{Sec.QuantumCosmology}Quantum Cosmology}
%%%%%%%%%%%%
%
% Quantum Cosmology
%
%%%%%%%%%%%
We first quantized the Friedmann equation, the key governing equation for cosmic evolution. We followed closely the procedure outlined in in Ref.~\cite{Bojowald_2015}. Throughout this paper, we took the normalization that $G = \hbar = 1$, where $G$ is the gravitational constant and $\hbar$ is the Planck constant divided by $2\pi$.

With the current observational supports, we already know that the curvature of universe is negligible so the Friedmann equation reads
\begin{align}
	\left(\frac{\dot{a}}{a}\right)^2 & = \frac{8\pi\rho}{3},
\end{align}
where $a$ is the scale factor quantifying the cosmic spacial evolution, $\rho$ is the average energy density of the universe at a given time, and the derivative is with respect to the physical time. This can be rewritten as 
\begin{align}
	-\frac{3}{8\pi}a\dot{a}^2 + a^3\rho = 0,
\end{align}
where the two terms can be regarded as the contribution from the space-time geometry and the energy of cosmic contents respectively.
To canonically quantize the equation, we first took the partial derivative of the action by $\dot{a}$ to obtain the momentum conjugate to the scale factor
\begin{align}
	p_a = -\frac{3}{4\pi}a\dot{a},
\end{align}
and the reparameterization generator
\begin{align}
	H = \frac{2 \pi}{3} p_a^2 + \rho a^4.
\end{align}
We note that $\rho \propto a^{-4}$ in the early universe so we define $\rho a^4=1/D$ where $D$ is a real constant.
Finally, similar to the classical canonical quantization, we could treat $p_a$ and $a$ as operators to obtain our constrained Hamiltonian equation
\begin{align}
	H\psi &= 0,
\end{align}
where
\begin{align}
	H &= \frac{2\pi}{3}\frac{\partial^2}{\partial a^2}+\frac{1}{D}.
\label{Eq.ConstrainedEq}
\end{align}
This equation has an apparent form for its solution 
\begin{align}
	\psi  \propto e^{\pm i \sqrt{\frac{3}{2\pi D}} a}. \label{Eq.ConstrainedEqSolution}
\end{align}
To conduct related computations on quantum computers, we replaced $\psi$ with $|\psi \rangle$ to obtain the bracket form 
\begin{align}
	H|\psi \rangle = 0 
	\Rightarrow \langle \psi |H|\psi \rangle =0.
\end{align}

\section{\label{Sec.Methodology}Methodology }
%%%%%%%%
%
%Methodology
%
%%%%%%%%%%%%

In this section, we first constructed a quantum gate representation for the expectation value $\langle \psi |H |\psi \rangle$. We then introduced our method for combining the strengths of both quantum and classical computers to find the solution for $\langle \psi |H |\psi \rangle=0$, by invoking the Probabilistic Bisection Algorithm (PBA). Finally we discussed how we mitigated the errors in implementing our algorithm on quantum computers.

With the Hamiltonian $H$ given earlier and $\psi (\theta) \propto e^{\pm i \theta a}$, we first considered how to evaluate
the eigenvalue $\langle \psi (\theta) |H(\theta) |\psi (\theta)\rangle $ on quantum computers. We discretized $H$ as
\begin{align}
	H = \frac{2 \pi}{3} \Delta^2 +\frac{1}{D},
	\label{eq.H_delta}
\end{align}
where $\Delta^2$ is the discrete version of the differential operator on an uniform grid. For a data sequence ${\bf v}\equiv\{v_j\}$ on a uniform grid of interval $\delta$, the discrete differential operation appears as
\begin{align}
\Delta^2 {\bf v} 
= \left\{  \frac{1}{\delta^2}\left(v_{j-1}-2v_j+v_{j+1}\right)  \right\}
 =  \frac{1}{\delta^2} \mathbf{M}_n{\bf v},
 \label{eq.delta2_v}
\end{align}
where $n$ is the number of qubits and $\mathbf{M}_n$ is a tridiagonal matrix of size $2^n \times 2^n$ in the form
\begin{align}
\mathbf{M}_n = 
	\begin{pmatrix}
		-2 &  1  	& 0    	&\hdots & 0\\
		1  & -2  	& 1    	&\ddots & \vdots\\
		0  &  1  	&\ddots    	& 1      & 0 \\
	\vdots & \ddots &1  	&-2      & 1\\
		0  & \hdots &0 		&1  	&-2
	\end{pmatrix}.
	\label{M_n}
\end{align}
We then employed a new Pauli-string representation of the matrix $\mathbf{M}_n$, as extended from the work in Ref.~\cite{Liu_2021} but much more efficient than their work. We left the details of this part in Appx.~\ref{App.CircuitRepresentationFD}.

In the implementation of $\Delta^2$, we devised a strategy that is a good tradeoff between efficiency and accuracy.
We noted that Ref.~\cite{Sato_2021} proposed an efficient method to calculate $\Delta^2$ but with the price of involving switch gates, which were known to be expensive in gate depth and would increase the noise. 
In our approach, we enforced the periodic boundary condition (PBC) for two advantages. First, it halved the number of terms to be calculated (see Appx.~\ref{App.CircuitRepresentationFD} for details). Secondly, it avoided the end-point problem in a finite-difference approach (i.e.~at $j=1$ or $j=2^n$ in Eq.~(\ref{eq.delta2_v})).
With the PBC, the eigenstate $\psi(\theta)$ takes the discrete form
\begin{align}
\psi_k (\theta) \propto e^{\pm i \theta a_k}
\end{align}
where
\begin{align}
a_k = k \delta = k \frac{2 \pi }{\theta 2^n} , \ k = 0, \cdots, 2^{n}-1.
\label{eq.a_k}
\end{align}
The fact that $\psi_k \propto e^{\pm i \frac{2 \pi k }{ 2^n}}$ is independent of $\theta$ is simply because the phase of eigenstate function spans over a $2\pi$ period regardless of the grid length and $\theta$.
On the other hand, $H$ carries a $\delta$ dependence. Substituting the form of $\delta$ in Eq.~(\ref{eq.a_k}) into Eqs.~(\ref{eq.H_delta}) and (\ref{eq.delta2_v}) gives
\begin{align}
H = \frac{\theta^2 2^{2n}}{6\pi}\mathbf{M}_n+\frac{1}{D}.
\end{align}
Therefore, the eigenvalue
$\langle \psi (\theta) |H(\theta) |\psi (\theta)\rangle$ is now simplified as 
$\langle \psi  |H(\theta) |\psi \rangle$.

We moved on to the gate representation of the initial eigenstate $|\psi\rangle$. First, similar to constructing the Quantum Fourier Transformation, we constructed a diagonal matrix
\begin{align}
\bigotimes_{l=1}^{n} \mathbf{P}\left(\frac{2\pi}{2^l}\right) = {\rm diag} \left(1,e^{i\frac{2\pi}{2^n}},e^{i\frac{4\pi}{2^{n}}}\cdots,e^{i\frac{(2^{n}-1)2\pi}{2^{n}}}\right),
\label{Eq.ExpDiagMatrix}
\end{align}
where $\mathbf{P}$ is the phase gate
\begin{align}
\mathbf{P}(\theta) = 
\begin{pmatrix}
1 & 0 \\
0 & e^{i\theta}
\end{pmatrix}.
\end{align}
Multiplying an uniform vector to this diagonal matrix would produce the correct form for $\psi$.
An uniform vector could be readily constructed by applying the Hadamard gate $\mathbf{H}$ to all qubits of zero state $|0\rangle_l$ ($l=1, ..., n$ representing the qubit index):
\begin{align}
 \bigotimes_{l=1}^{n} \mathbf{H}|0\rangle_l = 2^{-\frac{n}{2}}\sum_{k=0}^{2^{n}-1} |k\rangle.
\end{align}
Finally, our initial eigenstate $\psi$ could be constructed as 
\begin{align}
\left\{\psi_k\right\} & = \bigotimes_{l=1}^n \mathbf{P}\left(\frac{2\pi}{2^l}\right) \mathbf{H} |0\rangle_l \\
& = 2^{-\frac{n}{2}}\sum_{k=0}^{2^{n}-1} e^{\frac{2\pi i k}{2^n}}|k\rangle 
\end{align}
This completed our framework to compute $\langle \psi  |H(\theta) |\psi \rangle $.

Our ultimate goal is to find $\theta_0$ at which $\langle \psi  |H(\theta) |\psi \rangle |_{\theta = \theta_0} = 0$.
To this end, we invoked the PBA as illustrated in Ref.~\cite{Waeber2016}. 
It was an iterative process to locate  $\theta_0$.
It started off with a uniform prior within a reasonable domain $[\theta_a,\theta_b]$, where $\theta_a \leq \theta_0 \leq \theta_b$,
and bisectionally scaled up or down the probability density function (PDF) on the two sides of its median $\theta_{\rm med}$ based on the sign of the value 
$\langle \psi  |H(\theta_{\rm med}) |\psi \rangle$ 
in every iteration.
Eventually $\theta_{\rm med}$ would approach to $\theta_0$ with the desired accuracy.
In our current study, we have chosen $[\theta_a,\theta_b]=[0,2]$.
As compared with Ref.~\cite{Waeber2016}, the only difference in our work is that our bisectional scaling factor for the PDF was a constant while that in Ref.~\cite{Waeber2016} had a more sophisticated form that was computationally more expensive.
This makes our approach much more efficient without losing the key feature of PBA.

To combine the strengths of both classical and quantum computers,
we employed an HQC algorithm using the classical computers to determine $\theta_{\rm med}$ and update PDF
while using quantum computers to evaluate 
$\langle \psi  |H(\theta_{\rm med}) |\psi \rangle$.
The typical number of iterations was 100.
For the reference to be compared with the results from quantum computers, we also obtained the numerical result $\theta_0 \approx 0.709$ solely on classical computers.

We noted that in literature there are also gradient-based algorithms such as the SPSA, which is commonly used in VQE.
However, the PBA is better in our application for at least three reasons.
First, the PBA is computationally less expensive, because for each iteration only the value at one point is needed for PBA while at least two are required to evaluate a gradient. In the era of NISQ device, this is particularly important.
Secondly, gradient based algorithms are normally troubled when a gradient is equal to zero or infinity. In our application, the gradient at our solution $\theta_0$ is exactly zero. Lastly, and probably the most importantly, the gradient was indetermined at $\theta = \theta_0$.

Finally we mitigated the errors in calculating $\langle \psi |H(\theta) |\psi \rangle$ on quantum computers, known to be mainly from the readout noise at the level of few percents. 
We coped with the readout errors by using a deformation matrix $\mathbf{D}$
\begin{align}
\mathbf{D} =
	\begin{pmatrix}
		1-\eta_0 & \eta_1 \\
		\eta_0 & 1-\eta_1
	\end{pmatrix},
\end{align}
where
$\eta_0$ is the probability of measuring 0 as 1
and 
$\eta_1$ is the probability of measuring 1 as 0.
The values for $\eta_0$ and $\eta_1$ were readily provided by the IBM server. 
Therefore, the original probability could be estimated from the probability ${\bf p}$ output by a qubit as
\begin{align}
{\bf\tilde p} = \mathbf{D}^{-1}{\bf p},
\end{align}
where the negative components in ${\bf\tilde p}$, if any, were set to zeros.
For details about this correction process for a multi-qubit system, please refer to Appx.~\ref{Sec.ReadoutErrorCorrection}.

\begin{figure}[htbp]
	\centering 
	\includegraphics[width=\linewidth]{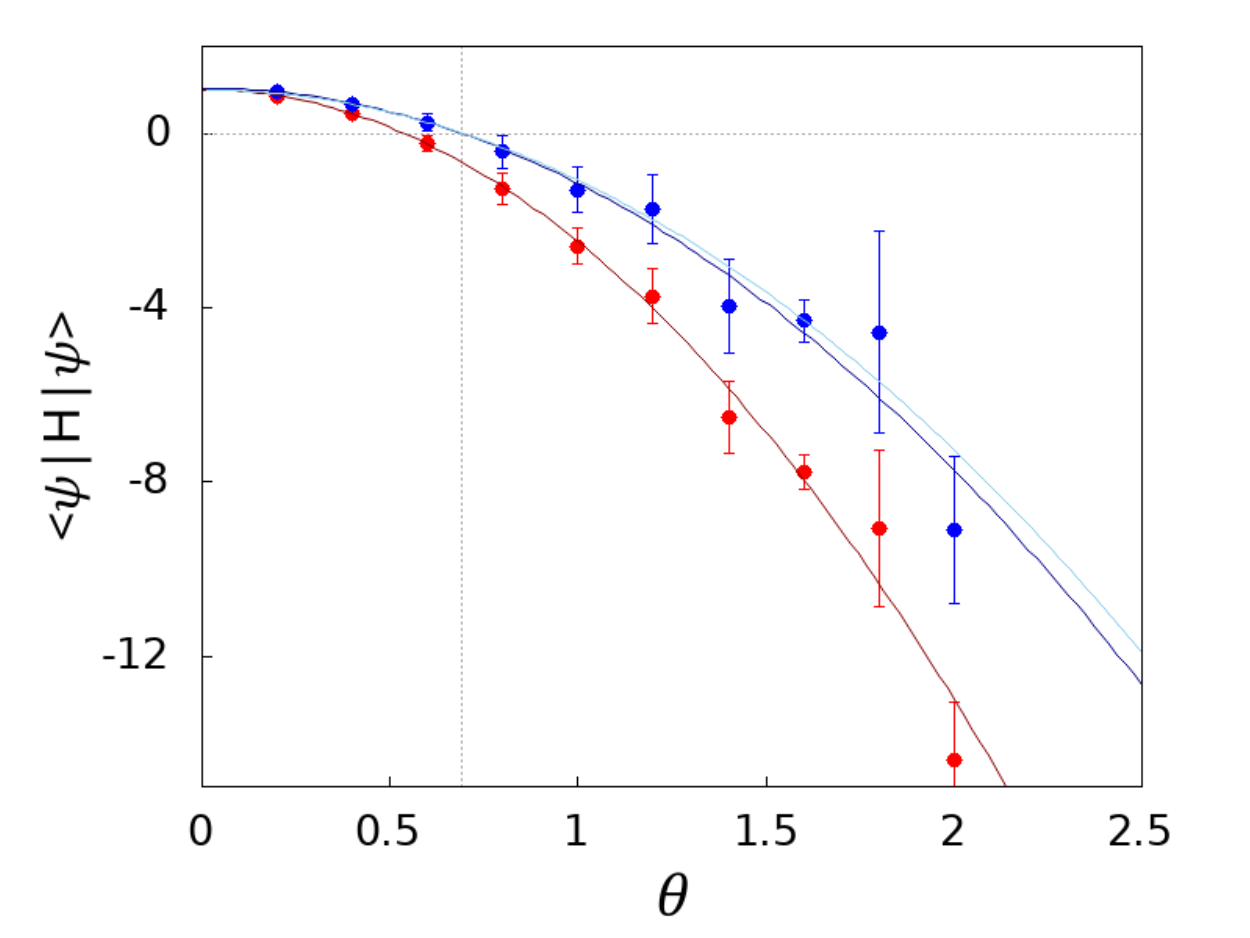}
	\caption{\justifying
		$\langle \psi| H(\theta) | \psi \rangle$ as a function of $\theta$ using the HQC algorithm based on the quantum processor ibmq\_manila, with (blue) or without (red) the readout-error correction, and their fitted curves. 
	The data points and 1-$\sigma$ error bars were estimated from four realizations.
	Also plotted for comparison is the numerical result (cyan) solely obtained from a classical computer. 
	The error in $\theta_0$ using the fitted curve (blue) for the corrected HQC result was only about $0.7\%$.}
	\label{Fig.theta_vs_pHp_Manila_allall_with_corr}
\end{figure}

To demonstrate the effectiveness of this error correction, we performed our method using three qubits in the quantum processor ibmq\_manila of IBM Canary.
A single execution of the quantum algorithm contained 15,000 shots, and we delivered 100 iterations for our HQC algorithm to solve for the constraint equation.
We repeated this process four times to quantify the statistical consistency of our results.
Fig.~\ref{Fig.theta_vs_pHp_Manila_allall_with_corr} shows the function $\langle \psi |H(\theta) |\psi \rangle $ with or without the readout-error correction, as compared with the numerical result solely obtained from a classical computer. It is clear that the correction was very effective, indicating that the errors in the outputs of quantum computers were indeed dominated by the readout noise.

%Figure1 <psi|H|psi>(theta)
% demonstration for readout error correction

\section{\label{Sec.ResultandDiscussion}Result and Discussion}
%%%%%%%%%%%%
%
%Result and Discussion
%
%%%%%%%%%%%%

\begin{figure}[htbp]
	\centering 
	\begin{subfigure}{0.32 \textwidth} 
		\includegraphics[width=\textwidth]{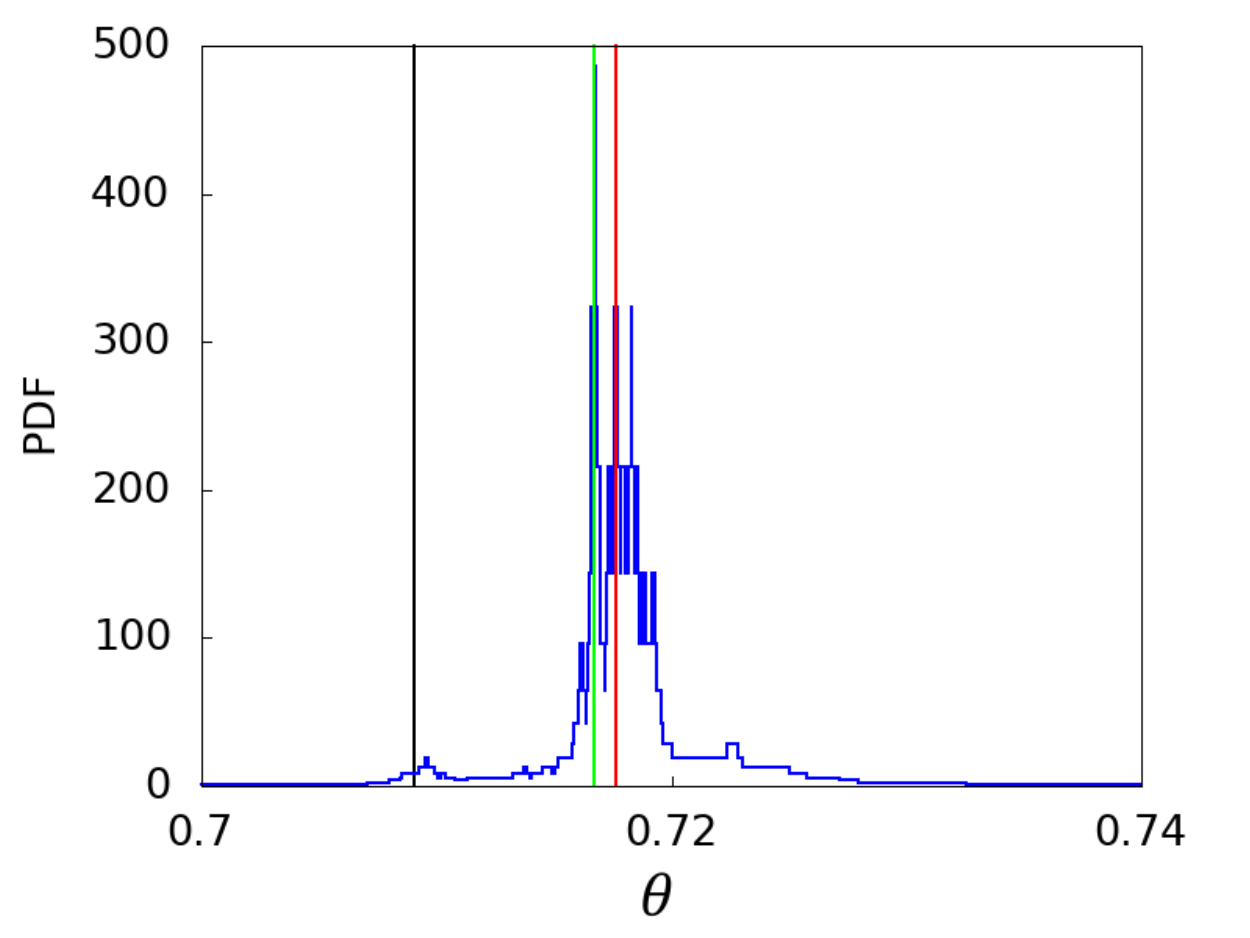} 
		\caption{} 
		\label{Fig.PBA_prob_density_lima_r4} 
	\end{subfigure} 
	\begin{subfigure}{0.32 \textwidth} 
		\includegraphics[width=\textwidth]{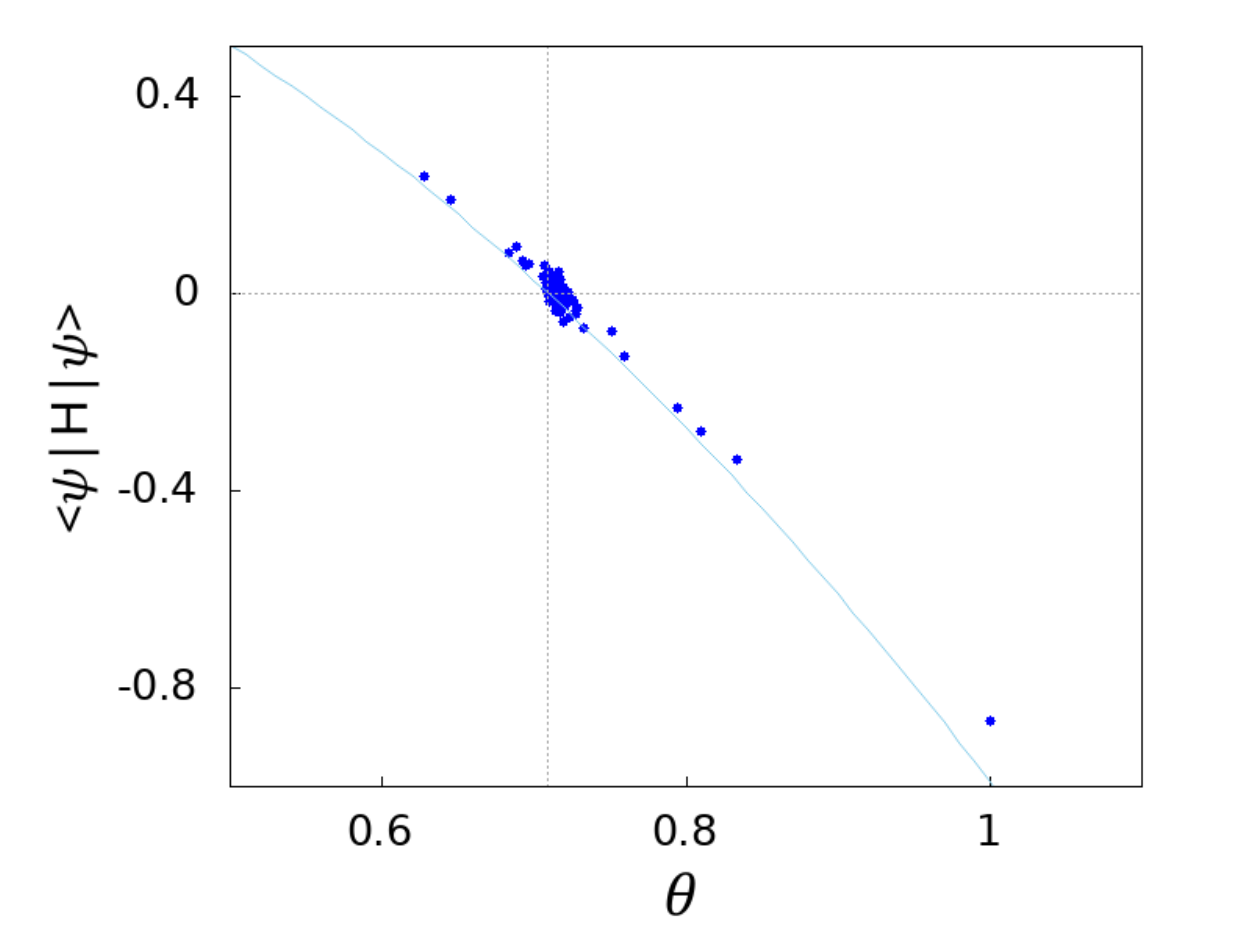} 
		\caption{} 
		\label{Fig.PBA_theta_vs_expH_lima_r4} 
	\end{subfigure} 
	\begin{subfigure}{0.32 \textwidth} 
		\includegraphics[width=\textwidth]{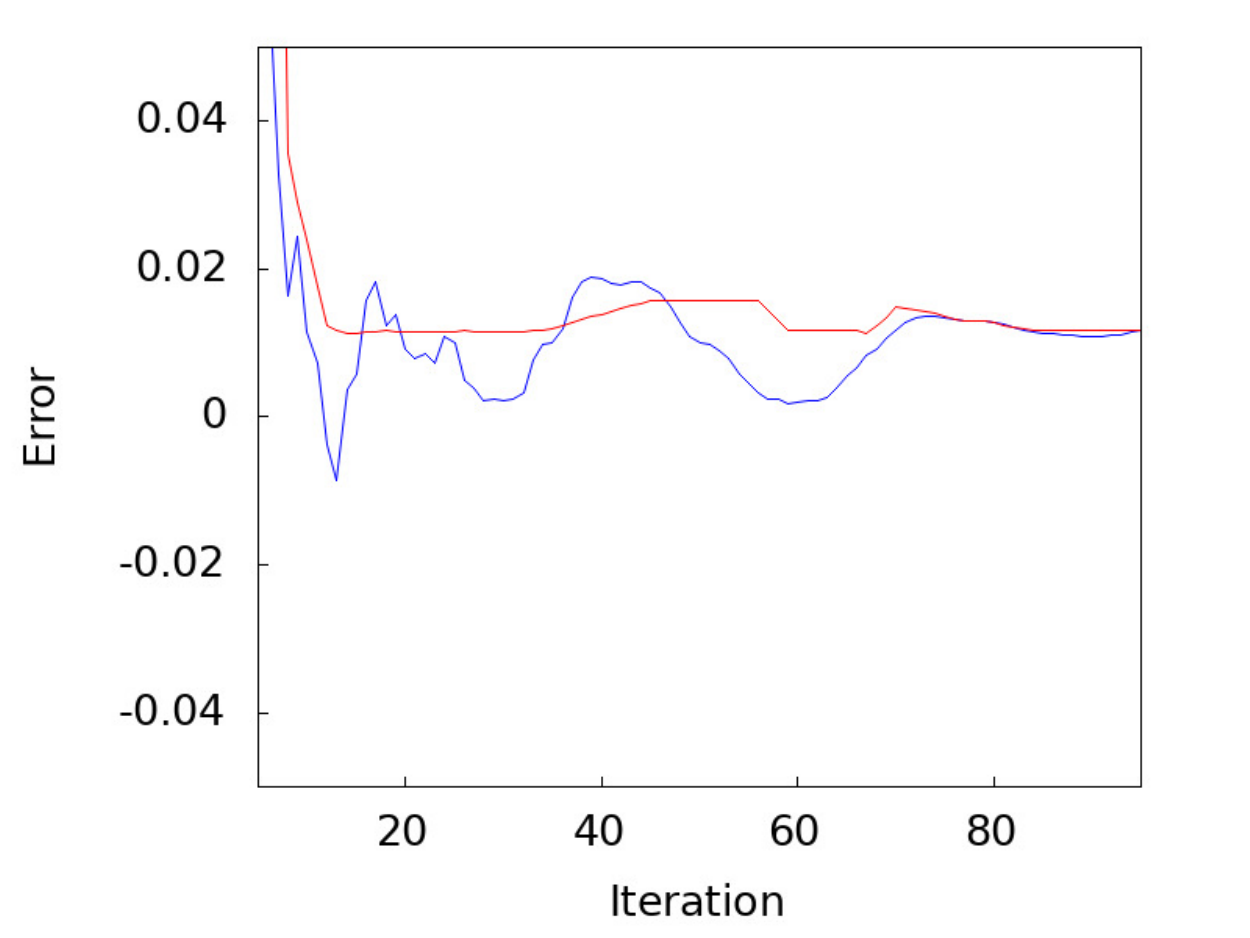} 
		\caption{} 
		\label{Fig.PBA_avg_iter_normal_lima_r4} 
	\end{subfigure}
	\caption{\justifying
		Demonstration of our HQC algorithm using four qubits in the quantum processor ibmq\_lima of IBM Canary. 100 iterations were implemented, each with 15,000 shots.
		(a) The final PDF after 100 iterations, and its median $\theta_{\rm med}$ (red), maximum $\theta_{\rm max}$ (green) and the true $\theta_0$ (black);
		(b) the eigenvalues $\langle \psi |H (\theta)|\psi \rangle $ for each of the 100 iterations (blue dots), as compared with the theoretical curve;
		(c) the fractional errors $\epsilon_{\theta_{\rm med}}$ (blue) and $\epsilon_{\theta_{\rm max}}$ (red) for each iteration.
		}
	\label{Fig.lima_r4}
\end{figure}

In this section, we presented our step-by-step results and discussed the origins and levels of the statistical and possible systematic errors in the results.
For quantum computing, we used four qubits in the quantum processor ibmq\_lima of IBM Canary.
Fig.~\ref{Fig.lima_r4} showed the results of a typical realization.
Fig.~\ref{Fig.PBA_prob_density_lima_r4} was the final PDF after 100 iterations. The median $\theta_{\rm med}$ (red) and the maximum $\theta_{\rm max}$ (green) were reasonably close to the true $\theta_0$ (black) within 1\% error.
Fig.~\ref{Fig.PBA_theta_vs_expH_lima_r4} showed the eigenvalues $\langle \psi |H (\theta)|\psi \rangle $ for each of the 100 iterations with different $\theta$, which converged to $\theta_0$. The solid curve was the numerical result obtained solely from a classical computer. Ideally, we hoped that all data points lied on the curve and converged to $\langle \psi |H (\theta_0)|\psi \rangle=0$. Therefore, the deviations of data points from the curve here indicated the errors. We observed that the estimated eigenvalues $\langle \psi |H (\theta)|\psi \rangle$ were systematically larger than the true values at the few-percent level.
Fig.~\ref{Fig.PBA_avg_iter_normal_lima_r4} showed the fractional errors $\epsilon_{\theta_{\rm med}}=\theta_{\rm med}/\theta_0 -1$ (blue) and $\epsilon_{\theta_{\rm max}}=\theta_{\rm max}/\theta_0 -1$ (red) for each iteration. It indicated the trend of convergence for the estimated $\theta_{\rm med}$ (and $\theta_{\rm max}$) to $\theta_0$ over the iterations. The oscillatory feature in the errors tended to disappear after 100 iterations.
For ease of this observation, we have convolved the results with a top-hat window of size 20.
In this realization, we also saw in Fig.~\ref{Fig.PBA_prob_density_lima_r4} and \ref{Fig.PBA_avg_iter_normal_lima_r4} that the final $\theta_{\rm med}$ was larger than $\theta_0$, likely due to the systematic overestimation for $\langle \psi |H (\theta)|\psi \rangle $. We would look into this in more details below.

%Figure2(a)(b)(c) theta vs. prob (error corrected, orig, fully correct)

\begin{figure}[!htbp]
	\centering 
	\includegraphics[width=0.5\textwidth]{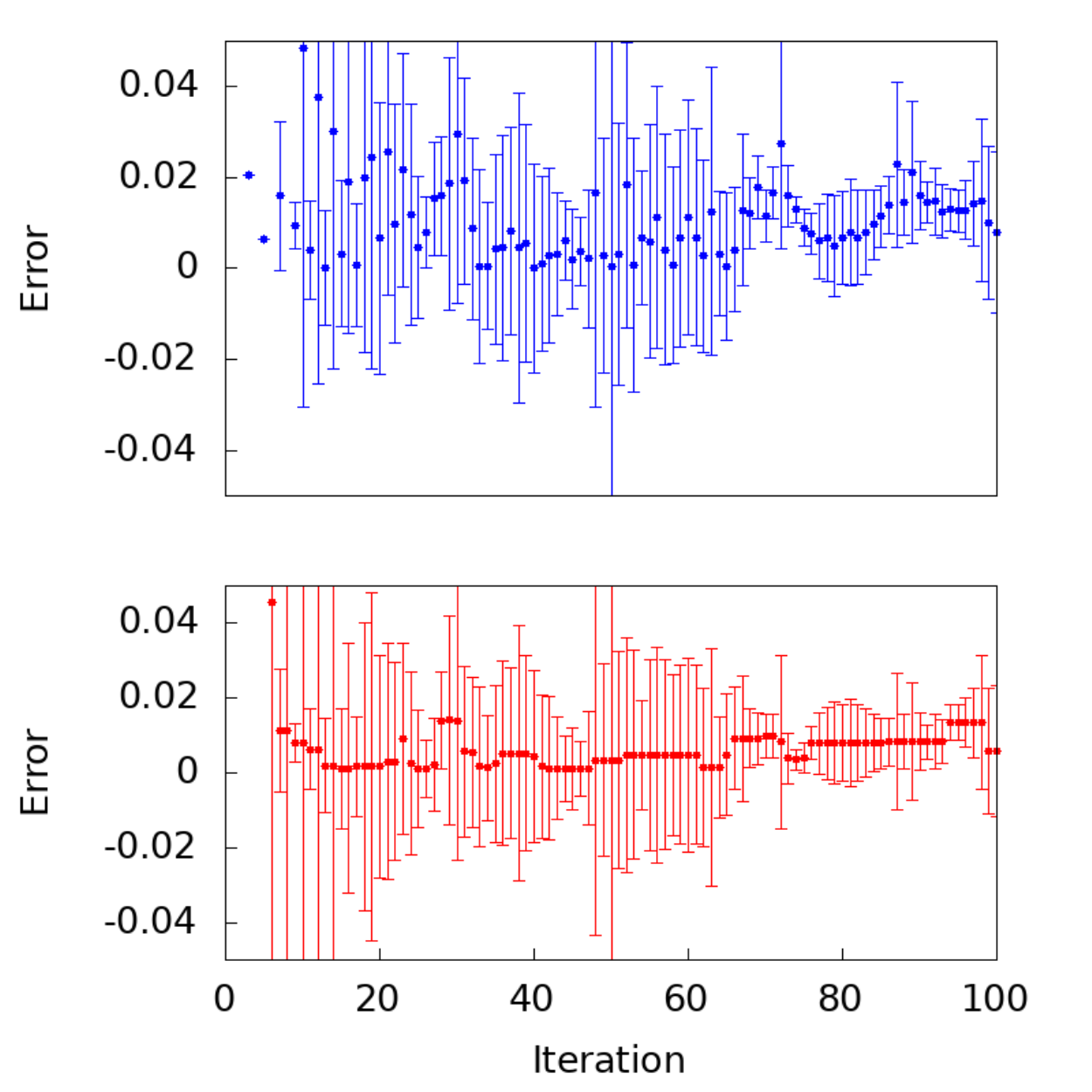} 
	\caption{\justifying
		The errors $\epsilon_{\theta_{\rm med}}$ (upper panel; blue) and $\epsilon_{\theta_{\rm max}}$ (lower panel; red) as a function of the iteration, averaged over the four realizations on ibmq\_lima.}
	\label{Fig.PBA_avg_over_normal_run}
\end{figure}

To study the statistical consistency of our results, we performed four realizations on the ibmq\_lima.
Fig.~\ref{Fig.PBA_avg_over_normal_run} showed the statistical results for $\epsilon_{\theta_{\rm med}}$ (upper panel; blue) and $\epsilon_{\theta_{\rm max}}$ (lower panel; red) inferred from the four runs.
The averages and the 1-$\sigma$ error bars for each iteration were calculated over the four realizations.
Again we observed the systematic overestimation of $\theta$, which were nevertheless well within the size of error bars.  
When we took the average of the last five iterations for ${\theta_{\rm max}}$ and ${\theta_{\rm med}}$, we obtained the percentage errors of about $1.0\%$ and $1.2\%$ respectively.

% Figure3 average of median and max, Avg over 4 runs

From Fig.~\ref{Fig.PBA_avg_iter_normal_lima_r4} and Fig.~\ref{Fig.PBA_avg_over_normal_run}, we observed that the convergence of $\theta$ to $\theta_0$ was effective after 100 iterations.
Therefore, instead of ending the iterative process by achieving certain preset accuracy, we chose to end our process after 100 iterations. This is a plausible tradeoff between the accuracy and the expenses of quantum computing. 100 iterations is large enough for showcasing the trend of convergence, yet short enough in computation time for obtaining good results.

\begin{figure}[!htbp]
	\centering 
	\includegraphics[width=0.45\textwidth]{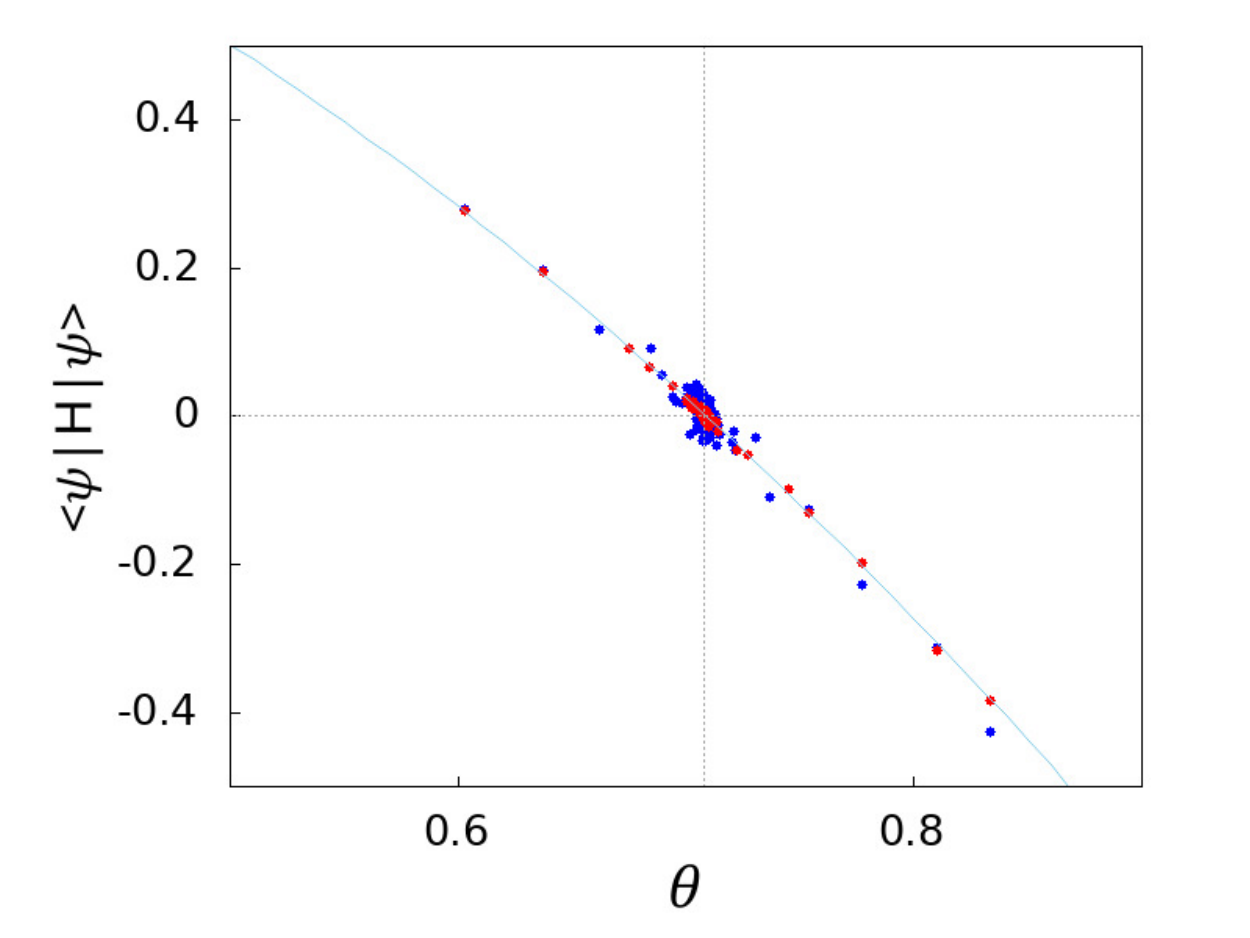} 
	\caption{\justifying
		Comparison of two realizations based on 15,000 shots (blue dots) and 200,000 shots (red dots) in each iteration. The results of 200,000-shot based realization scattered much less from the theoretical curve (cyan).}
	\label{Fig.PBA_theta_vs_expH_compare_shots}
\end{figure}

To further understand the properties of our errors, we performed two realizations of our algorithm on the quantum simulators, each with 100 iterations of 15,000 or 200,000 shots. Fig.~\ref{Fig.PBA_theta_vs_expH_compare_shots} showed the comparison.
It was clear that the results based on 200,000 shots for each iteration scattered much less from the theoretical curve, indicating less errors in the estimated $\langle \psi |H (\theta)|\psi \rangle $. Specifically, the Root Mean Square (RMS) values for the 200,000- and 15,000-shot based realizations are about 175 and 41 respectively, roughly in line with the statistical scalability for the number of shots.
We note that the quantum computation time is proportional to the number of shots $n_{\rm s}$ 
while the statistical error is inversely proportional to $\sqrt{n_{\rm s}}$.
This means that as compared with a $15,000$-shot realization, a $200,000$-shot realization will take about 13 times more time while improving the statistical error only by a factor of less than 4. This further signals the importance of having a reasonable tradeoff between the accuracy and the quantum computation expenses, as already emphasized earlier with slightly different concerns but the same goal.
The typical computation time for a $15,000$-shot based realization using our HQC algorithm to solve for the constraint equation studied here is about one hour.
 
% 4 Comparison between 15000 and 200,000 shots on simulation

Tab.~\ref{Tab.ResutSimQC} summarizes the performance of our HQC algorithm as discussed above but in terms of the errors in $\langle \psi|H(\theta) |\psi \rangle \equiv \langle H(\theta)\rangle$ and $\theta_0$.
We define the error in $\langle H(\theta)\rangle$ as $\epsilon_{\langle H\rangle}(\theta) = \langle H(\theta) \rangle_{\rm QC} - \langle H(\theta) \rangle_{\rm thy}$, where $\langle H(\theta) \rangle_{\rm QC}$ is the value estimated by each iteration in our HQC algorithm and $\langle H(\theta) \rangle_{\rm thy}$ is the theoretical value of the same $\theta$ calculated solely by a classical computer.
The second and third columns in the table are the average errors $\mu_{\epsilon_{\langle H\rangle}}=\langle \epsilon_{\langle H\rangle}(\theta) \rangle_\theta$ and their standard deviations $\sigma_{\epsilon_{\langle H\rangle}}$, respectively, obtained over the entire 100 iterations of each realization.
The last column shows the fractional errors $\epsilon_{\theta_{\rm med}}$ in the final estimated $\theta_{\rm med}$ (after the 100 iterations) for each realization.

Eight realizations obtained from simulators or physical quantum computers are summarized here.
The first four rows are the results from quantum simulators, with or without the simulated noise, and with or without the readout error correction. These are consistent with the earlier discussions associated with Fig.~\ref{Fig.theta_vs_pHp_Manila_allall_with_corr} and Fig.~\ref{Fig.PBA_theta_vs_expH_compare_shots}, but provide more quantitative details.
They emphasize the statistical dependence of accuracy on the number of shots (in the comparison between the first and second rows) and the effectiveness of our method for the readout error correction (in the comparison among the first, third, and fourth rows).
The last four rows are the results from the quantum processor ibmq\_lima, all with readout error correction.
Again we saw the seemingly systematic overestimation in the final estimated $\theta_{\rm med}$ though at the level of only 1\%.

It is noteworthy that the errors in the final $\theta_{\rm med}$ estimated by the physical quantum computer (the last four rows) are nearly an order of magnitude higher than those from the simulators (the first and fourth rows). This disparity reflects the current quality of the quantum computer in use and should be considered when relying solely on simulators to explore any research topic.

% Error table
\begin{table*}[!htbp]
	\caption{\justifying
		\label{Tab.ResutSimQC}
		Comparison of errors in $\langle \psi|H |\psi \rangle \equiv \langle H\rangle$ and $\theta_0$ for results obtained from simulators and physical quantum computers.
		Numbers in the second and third columns are the average errors $\mu_{\epsilon_{\langle H\rangle}}$ and their standard deviations $\sigma_{\epsilon_{\langle H\rangle}}$, respectively, obtained over the entire 100 iterations of each realization.
		The last column shows the fractional errors $\epsilon_{\theta_{\rm med}}$ in the final estimated $\theta_{\rm med}$ for each realization.
 }
	\begin{ruledtabular}
		\begin{tabular}{lccc}
			& $\mu_{\epsilon_{\langle H\rangle}}$ ($\times 10^{-3}$) & $\sigma_{\epsilon_{\langle H\rangle}}$ ($\times 10^{-3}$) & $\epsilon_{\theta_{\rm med}}$ ($\times 10^{-4}$) \\
			\hline 
			simulator results without noise, 15,000 shots & $ -3.3$ & 18 & $-27$ \\
			simulator results without noise, 200,000 shots & $ -0.78$ & 4 & $-14$ \\
			simulator results with noise, 15,000 shots & $-130$ & $23$ & $-657$ \\
			simulator results with noise $+$ error corr., 15,000 shots &  $-4.2$ & 17 & $-29$\\
			ibmq\_lima realization 1 $+$ error corr., 15,000 shots & $22$ & 35 & 220 \\
			ibmq\_lima realization 2 $+$ error corr., 15,000 shots & $-11$ & 63 & 181 \\
			ibmq\_lima realization 3 $+$ error corr., 15,000 shots & $-24$ & 71 & 2 \\
			ibmq\_lima realization 4 $+$ error corr., 15,000 shots & $22$ & 20 & 107
		\end{tabular}
	\end{ruledtabular}
\end{table*}

\section{\label{Sec.Conclusion}Conclusion}
%%%%%%%%%%
%
%Conclusion
%
%%%%%%%%
In this paper, we employed a newly proposed HQC algorithm, instead of the commonly used VQE, to address a key problem in Quantum Cosmology on physical quantum computers. This approach leverages the strengths of quantum computers to calculate eigenvalues and classical computers to handle the probability density function within the PBA. After 100 iterations, each consisting of 15,000 quantum computing shots, we achieved an accuracy of approximately 1\% error in the solution. The computation time for a single realization of this exercise using four qubits is roughly one hour. We anticipate that further iterations and additional qubits will improve both accuracy and efficiency. This study serves as the first demonstration that quantum cosmology can be effectively explored on real quantum computers, with future advancements likely to benefit from an increase in qubit count and reductions in noise.

This sheds light on the future applications of quantum computers not only in quantum cosmology but also in quantum gravity in general. 
Solving constrained equations is often non-trivial; however, these equations can be reformulated as expectation value problems, i.e., $H\psi \rightarrow \langle \psi |H |\psi \rangle$. Therefore, quantum computers can significantly alleviate the challenges associated with these problems, as they are considerably more efficient at solving expectation value problems compared to classical computers. 
A further application of such an approach is to find the eigenstate of a complex Hamiltonian.
It was demonstrated in Ref.~\cite{Barends_2016} that a complex Hamiltonian $H$ could the end result of an adiabatic process starting from a simple Hamiltonian, and the corresponding end eigenstate $\psi$ could be found by evolving the eigenstate of the initial simple Hamiltonian through the same adiabatic process. Therefore, in principle we could use quantum computers to first find the initial eigenstate and then evolve it through the adiabatic process to find $\psi$. We shall demonstrate this in our future work.
Such capability positions quantum computers in an even more powerful role for addressing problems that are typically intractable for classical computers.

\begin{acknowledgments}
This research used the IBM Quantum services. Our views expressed here do not reflect the official policy or position of IBM or the IBM Quantum team.
We acknowledge the support from National Science and Technology Council (NSTC), Taiwan.
\end{acknowledgments}

\appendix
\section{An Efficient Realization of $\frac{d^2}{d x^2}$ using Finite Difference on Quantum Computers}
\label{App.CircuitRepresentationFD}

Here, we present a Pauli-string representation of the finite difference matrix $\mathbf{M}_n$, building upon the work in Ref.~\cite{Liu_2021} with significantly improved efficiency.
As demonstrated in Ref.~\cite{Liu_2021}, dividing the tridiagonal matrix $\mathbf{M}_n$ of size $2^n \times 2^n$ in Eq.~(\ref{M_n}) into four quadrants gives the form
\begin{align}
	\mathbf{M}_n = 
	\begin{pmatrix}
	\begin{array}{c|c}
		\mathbf{M}_{n-1}  & \mathbf{D}_{n-1}\\
		\hline
		\mathbf{D}_{n-1}^{\rm T} & \mathbf{M}_{n-1}	 
	\end{array}
	\end{pmatrix},
	\label{M_n_A}
\end{align}
where
\begin{align}
	\mathbf{D}_{n-1} = 
	\begin{pmatrix}
		\begin{array}{ccccc}
			0&0&\hdots&0&0	\\
			0&0&\hdots&0&0	\\
			\vdots&\vdots&\ddots&0&\vdots\\
			0&0&\hdots&0&0	\\
			1&0&\hdots&0&0	 	 
		\end{array}
	\end{pmatrix}.
\end{align}
The $\mathbf{D}_{n-1}$ and $\mathbf{D}_{n-1}^{\rm T}$ can then be constructed from the Pauli matrices $\sigma_1$ and $\sigma_2$ by taking a linear sum of their tensor products:
\begin{align}
	\mathbf{D}_{n-1} & = \otimes_{m=1}^{n-1}\sigma_{-}  \label{Eq.DmRepresentation},\\
	\mathbf{D}^T_{n-1} & = \otimes_{m=1}^{n-1}\sigma_{+}  \label{Eq.DmTRepresentation},
\end{align}
where $\sigma_\pm = \frac{1}{2} (\sigma_1\pm i \sigma_2)$.
This means that the construction of $\mathbf{M}_n$ is a linear combination of tensor products of the simple operators $\{ {\bf I},\sigma_{+},\sigma_{-}\}$, starting iteratively from $n=1$ using Eq.~(\ref{M_n_A}).
Therefore, the total number of terms to be computed for constructing $\mathbf{M}_n$ is $\sum_{m=1}^n 2^{m+1}$, where $n$ is the number of qubits

To improve the efficiency for constructing $\mathbf{M}_n$, we propose a method to reduce the required number of terms by a factor of more than five. It involves three considerations. 
The first is to notice that $\mathbf{D}_{n-1}$ and $\mathbf{D}^T_{n-1}$ are the first and third quadrants, respectively, of a Hermitian matrix:
\begin{align}
\mathbf{D}_{n-1{\rm (t)}} = 
\begin{pmatrix} 
	\begin{array}{cccc|cccc}
		0&\hdots&0&0&0&0&\hdots&0\\ 
		\vdots&\ddots&\vdots&\vdots&\vdots&\vdots&\iddots&\vdots\\ 
		0&\hdots&0&0&0&0&\hdots&0\\ 
		0&\hdots&0&0&1&0&\hdots&0\\ 
		\hline
		0&\hdots&0&1&0&0&\hdots&0\\ 
		0&\hdots&0&0&0&0&\hdots&0\\ 
		\vdots&\iddots&\vdots&\vdots&\vdots&\vdots&\ddots&\vdots\\ 
		0&\hdots&0&0&0&0&\hdots&0 
	\end{array}
\end{pmatrix}.
\end{align}
Because $\mathbf{D}_{n-1}$ and $\mathbf{D}^{\rm T}_{n-1}$ are now combined into a single Hermitian matrix $\mathbf{D}_{n-1{\rm (t)}} $, it halves the number of terms to be computed.

Secondly, the terms involving an odd number of $\sigma_2$ in a tensor product would be purely imaginary so their sum has to vanish, allowing for us to completely ignore them.
Specifically, the number of residual terms is $\sum_{l=1}^n \sum_{m=0}^M C_{2m}^l$, where $M$ is the maximum integer such that $2M \leq l$. 
This equals to $\sum_{l=1}^n 2^{l-1}$, providing another halving in the number of terms.
For example, for $n = 2, 3, 4$, the numbers of necessary terms are now reduced from the original $12, 28, 60$ down to $3, 7, 15$.

Our last consideration is to set the tridiagonal matrix $\mathbf{M}_n$ to be periodic at its boundaries. This is easily achieved by setting the components at the upper right and bottom left corners to one. Under this periodic boundary condition, all terms with $\sigma_2$ being the leading term in the tensor products for the last iteration in constructing $\mathbf{M}_n$ would also vanish. This halves the number of terms in the last iteration, so the total number of terms is now $\sum_{l=1}^{n-1} \sum_{m=0}^M C_{2m}^l + \frac{1}{2}\sum_{m=0}^M C_{2m}^n$.
For example, for $n = 2, 3, 4$, the numbers of required terms are further reduced to $2, 5, 11$.
This provides an extra reduction factor of $1.5$ for $n=2$, which asymptotically approaches $4/3$ for large $n$.

Therefore, in total we have reduced the number of terms in constructing $\mathbf{M}_n$ by a factor of $2\times 2\times 1.5 =6$ for $n=2$, which asymptotically approaches $2\times 2\times 4/3  \approx 5.3$ for large $n$.
Such improvement in efficiency would dramatically save the quantum computation time.

\section{Readout Error Correction}\label{Sec.ReadoutErrorCorrection}
%%%%%%%%%%%%%%%%%
% 
% Readout correction
%
%%%%%%%%%%%%%%%%%%

Here, we demonstrate how the readout error could be corrected in a multi-qubit system.
First we construct a deformation matrix to correct for the readout error of a single qubit:
\begin{align}
\mathbf{D} =
	\begin{pmatrix}
		1-\eta_0 & \eta_1 \\
		\eta_0 & 1-\eta_1
	\end{pmatrix},
	\label{D}
\end{align}
where $\eta_1$ is the probability of measuring 1 as 0, and $\eta_0$ is the probability of measuring 0 as 1.
Therefore, for the probability data vector ${\bf p} = (p_0, \ p_1)$ output from a given qubit, the original probabilities can be estimated as
\begin{align}
	{\bf\tilde p} = \mathbf{D}^{-1}{\bf p}.
\end{align}
This can be generalized for a multi-qubit system in the form:
\begin{align}
	{\bf\tilde P} = \mathbf{D}^{-1}_{\rm all} {\bf P},
\end{align}
where ${\bf P}=\left\{P_{(j)} \right\}$ with $j=1, \dots, 2^n$ being the index of all readout channels in the $n$-qubit system,
${\bf\tilde P}$ is the data vector consisting of the estimated real probabilities, and $\mathbf{D}^{-1}_{\rm all}$ is an operation consisting of the following three steps.
First, let ${\bf s}$ be the data vector consisting of the integer outputs from all readout channels of the quantum computation,
then the estimated real values would be 
\begin{align}
	{\bf\tilde s}_{\rm o} = \left[\bigotimes_{i=1}^n \mathbf{D}^{-1}_{(i)}\right] {\bf s},
\end{align}
where 
$\mathbf{D}^{-1}_{(i)}$ is the deformation matrix in the form of Eq.~(\ref{D}) for the $i$-th qubit with components $\eta_{0(i)}$ and $\eta_{1(i)}$ that can be retrieved from the IBM server through the keywords
`Prob meas0 prep1' and `Prob meas1 prep0' respectively.
Normally ${\bf\tilde s}_{\rm o}$ carries non-integers and even negative components, so we further correct these with
\begin{align}
{\tilde s}_{(j)} = 
	\begin{cases}
		0, & \text{ if } {\tilde s}_{{\rm o}(j)} \leq 0, \\
		{\rm int}({\tilde s}_{{\rm o}(j)}), & \text{ if } {\tilde s}_{{\rm o}(j)} > 0,
	\end{cases}
\end{align}
where 
${\tilde s}_{\rm (j)}$ is the component of the vector ${\bf\tilde s}$ and 
${\rm int}(x)$ is an operation to round down $x$ to the nearest integer.
Finally the estimated probabilities with readout error correction can be obtained by 
\begin{align}
	{\bf\tilde P} = \frac{\bf\tilde s}{\sum_{j=1}^{2^n} {\tilde s}_{(j)}}.
\end{align}

\bibliography{QuantumCosmologyonQuantumComputerv2}% Produces the bibliography via BibTeX.

\providecommand{\noopsort}[1]{}\providecommand{\singleletter}[1]{#1}%
\begin{thebibliography}{18}%
\makeatletter
\providecommand \@ifxundefined [1]{%
 \@ifx{#1\undefined}
}%
\providecommand \@ifnum [1]{%
 \ifnum #1\expandafter \@firstoftwo
 \else \expandafter \@secondoftwo
 \fi
}%
\providecommand \@ifx [1]{%
 \ifx #1\expandafter \@firstoftwo
 \else \expandafter \@secondoftwo
 \fi
}%
\providecommand \natexlab [1]{#1}%
\providecommand \enquote  [1]{``#1''}%
\providecommand \bibnamefont  [1]{#1}%
\providecommand \bibfnamefont [1]{#1}%
\providecommand \citenamefont [1]{#1}%
\providecommand \href@noop [0]{\@secondoftwo}%
\providecommand \href [0]{\begingroup \@sanitize@url \@href}%
\providecommand \@href[1]{\@@startlink{#1}\@@href}%
\providecommand \@@href[1]{\endgroup#1\@@endlink}%
\providecommand \@sanitize@url [0]{\catcode `\\12\catcode `\$12\catcode
  `\&12\catcode `\#12\catcode `\^12\catcode `\_12\catcode `\%12\relax}%
\providecommand \@@startlink[1]{}%
\providecommand \@@endlink[0]{}%
\providecommand \url  [0]{\begingroup\@sanitize@url \@url }%
\providecommand \@url [1]{\endgroup\@href {#1}{\urlprefix }}%
\providecommand \urlprefix  [0]{URL }%
\providecommand \Eprint [0]{\href }%
\providecommand \doibase [0]{https://doi.org/}%
\providecommand \selectlanguage [0]{\@gobble}%
\providecommand \bibinfo  [0]{\@secondoftwo}%
\providecommand \bibfield  [0]{\@secondoftwo}%
\providecommand \translation [1]{[#1]}%
\providecommand \BibitemOpen [0]{}%
\providecommand \bibitemStop [0]{}%
\providecommand \bibitemNoStop [0]{.\EOS\space}%
\providecommand \EOS [0]{\spacefactor3000\relax}%
\providecommand \BibitemShut  [1]{\csname bibitem#1\endcsname}%
\let\auto@bib@innerbib\@empty
%</preamble>
\bibitem [{\citenamefont {Feynman}(1985)}]{Feynman:85}%
  \BibitemOpen
  \bibfield  {author} {\bibinfo {author} {\bibfnamefont {R.~P.}\ \bibnamefont
  {Feynman}},\ }\bibfield  {title} {\bibinfo {title} {Quantum mechanical
  computers},\ }\href {https://doi.org/10.1364/ON.11.2.000011} {\bibfield
  {journal} {\bibinfo  {journal} {Optics News}\ }\textbf {\bibinfo {volume}
  {11}},\ \bibinfo {pages} {11} (\bibinfo {year} {1985})}\BibitemShut {NoStop}%
\bibitem [{\citenamefont {{Deutsch}}\ and\ \citenamefont
  {{Jozsa}}(1992)}]{1992RSPSA.439..553D}%
  \BibitemOpen
  \bibfield  {author} {\bibinfo {author} {\bibfnamefont {D.}~\bibnamefont
  {{Deutsch}}}\ and\ \bibinfo {author} {\bibfnamefont {R.}~\bibnamefont
  {{Jozsa}}},\ }\bibfield  {title} {\bibinfo {title} {{Rapid Solution of
  Problems by Quantum Computation}},\ }\href
  {https://doi.org/10.1098/rspa.1992.0167} {\bibfield  {journal} {\bibinfo
  {journal} {Proceedings of the Royal Society of London Series A}\ }\textbf
  {\bibinfo {volume} {439}},\ \bibinfo {pages} {553} (\bibinfo {year}
  {1992})}\BibitemShut {NoStop}%
\bibitem [{\citenamefont {Kitaev}(1995)}]{kitaev1995quantum}%
  \BibitemOpen
  \bibfield  {author} {\bibinfo {author} {\bibfnamefont {A.~Y.}\ \bibnamefont
  {Kitaev}},\ }\href@noop {} {\bibinfo {title} {Quantum measurements and the
  abelian stabilizer problem}} (\bibinfo {year} {1995}),\ \Eprint
  {https://arxiv.org/abs/quant-ph/9511026} {arXiv:quant-ph/9511026 [quant-ph]}
  \BibitemShut {NoStop}%
\bibitem [{\citenamefont {Shor}(1997)}]{Shor_1997}%
  \BibitemOpen
  \bibfield  {author} {\bibinfo {author} {\bibfnamefont {P.~W.}\ \bibnamefont
  {Shor}},\ }\bibfield  {title} {\bibinfo {title} {Polynomial-time algorithms
  for prime factorization and discrete logarithms on a quantum computer},\
  }\href {https://doi.org/10.1137/s0097539795293172} {\bibfield  {journal}
  {\bibinfo  {journal} {{SIAM} Journal on Computing}\ }\textbf {\bibinfo
  {volume} {26}},\ \bibinfo {pages} {1484} (\bibinfo {year}
  {1997})}\BibitemShut {NoStop}%
\bibitem [{\citenamefont {Grover}(1996)}]{grover1996fast}%
  \BibitemOpen
  \bibfield  {author} {\bibinfo {author} {\bibfnamefont {L.~K.}\ \bibnamefont
  {Grover}},\ }\bibfield  {title} {\bibinfo {title} {A fast quantum mechanical
  algorithm for database search},\ }\href@noop {} {\bibfield  {journal}
  {\bibinfo  {journal} {Proceedings, 28th Annual ACM Symposium on the Theory of
  Computing (STOC)}\ ,\ \bibinfo {pages} {212}} (\bibinfo {year} {1996})},\
  \Eprint {https://arxiv.org/abs/quant-ph/9605043} {arXiv:quant-ph/9605043
  [quant-ph]} \BibitemShut {NoStop}%
\bibitem [{\citenamefont {Peruzzo}\ \emph {et~al.}(2014)\citenamefont
  {Peruzzo}, \citenamefont {McClean}, \citenamefont {Shadbolt}, \citenamefont
  {Yung}, \citenamefont {Zhou}, \citenamefont {Love}, \citenamefont
  {Aspuru-Guzik},\ and\ \citenamefont {O'Brien}}]{osti_1623945}%
  \BibitemOpen
  \bibfield  {author} {\bibinfo {author} {\bibfnamefont {A.}~\bibnamefont
  {Peruzzo}}, \bibinfo {author} {\bibfnamefont {J.}~\bibnamefont {McClean}},
  \bibinfo {author} {\bibfnamefont {P.}~\bibnamefont {Shadbolt}}, \bibinfo
  {author} {\bibfnamefont {M.-H.}\ \bibnamefont {Yung}}, \bibinfo {author}
  {\bibfnamefont {X.-Q.}\ \bibnamefont {Zhou}}, \bibinfo {author}
  {\bibfnamefont {P.~J.}\ \bibnamefont {Love}}, \bibinfo {author}
  {\bibfnamefont {A.}~\bibnamefont {Aspuru-Guzik}},\ and\ \bibinfo {author}
  {\bibfnamefont {J.~L.}\ \bibnamefont {O'Brien}},\ }\bibfield  {title}
  {\bibinfo {title} {A variational eigenvalue solver on a photonic quantum
  processor},\ }\bibfield  {journal} {\bibinfo  {journal} {Nature
  Communications}\ }\textbf {\bibinfo {volume} {5}},\ \href
  {https://doi.org/10.1038/ncomms5213} {10.1038/ncomms5213} (\bibinfo {year}
  {2014})\BibitemShut {NoStop}%
\bibitem [{\citenamefont {McClean}\ \emph {et~al.}(2016)\citenamefont
  {McClean}, \citenamefont {Romero}, \citenamefont {Babbush},\ and\
  \citenamefont {Aspuru-Guzik}}]{McClean_2016}%
  \BibitemOpen
  \bibfield  {author} {\bibinfo {author} {\bibfnamefont {J.~R.}\ \bibnamefont
  {McClean}}, \bibinfo {author} {\bibfnamefont {J.}~\bibnamefont {Romero}},
  \bibinfo {author} {\bibfnamefont {R.}~\bibnamefont {Babbush}},\ and\ \bibinfo
  {author} {\bibfnamefont {A.}~\bibnamefont {Aspuru-Guzik}},\ }\bibfield
  {title} {\bibinfo {title} {The theory of variational hybrid quantum-classical
  algorithms},\ }\href {https://doi.org/10.1088/1367-2630/18/2/023023}
  {\bibfield  {journal} {\bibinfo  {journal} {New Journal of Physics}\ }\textbf
  {\bibinfo {volume} {18}},\ \bibinfo {pages} {023023} (\bibinfo {year}
  {2016})}\BibitemShut {NoStop}%
\bibitem [{\citenamefont {Romero}\ \emph {et~al.}(2018)\citenamefont {Romero}
  \emph {et~al.}}]{Romero_2018}%
  \BibitemOpen
  \bibfield  {author} {\bibinfo {author} {\bibfnamefont {J.}~\bibnamefont
  {Romero}} \emph {et~al.},\ }\bibfield  {title} {\bibinfo {title} {Strategies
  for quantum computing molecular energies using the unitary coupled cluster
  ansatz},\ }\href {https://doi.org/10.1088/2058-9565/aad3e4} {\bibfield
  {journal} {\bibinfo  {journal} {Quantum Science and Technology}\ }\textbf
  {\bibinfo {volume} {4}},\ \bibinfo {pages} {014008} (\bibinfo {year}
  {2018})}\BibitemShut {NoStop}%
\bibitem [{\citenamefont {Arute}\ \emph {et~al.}(2020)\citenamefont {Arute}
  \emph {et~al.}}]{Google2020}%
  \BibitemOpen
  \bibfield  {author} {\bibinfo {author} {\bibfnamefont {F.}~\bibnamefont
  {Arute}} \emph {et~al.},\ }\bibfield  {title} {\bibinfo {title} {Hartree-fock
  on a superconducting qubit quantum computer},\ }\href
  {https://doi.org/10.1126/science.abb9811} {\bibfield  {journal} {\bibinfo
  {journal} {Science}\ }\textbf {\bibinfo {volume} {369}},\ \bibinfo {pages}
  {1084} (\bibinfo {year} {2020})}\BibitemShut {NoStop}%
\bibitem [{\citenamefont {Kandala}\ \emph {et~al.}(2017)\citenamefont {Kandala}
  \emph {et~al.}}]{IBM2017}%
  \BibitemOpen
  \bibfield  {author} {\bibinfo {author} {\bibfnamefont {A.}~\bibnamefont
  {Kandala}} \emph {et~al.},\ }\bibfield  {title} {\bibinfo {title}
  {{Hardware-efficient variational quantum eigensolver for small molecules and
  quantum magnets}},\ }\href {https://doi.org/10.1038/nature23879} {\bibfield
  {journal} {\bibinfo  {journal} {Nature}\ }\textbf {\bibinfo {volume} {549}},\
  \bibinfo {pages} {242} (\bibinfo {year} {2017})}\BibitemShut {NoStop}%
\bibitem [{\citenamefont {O'Malley}\ \emph {et~al.}(2016)\citenamefont
  {O'Malley} \emph {et~al.}}]{PhysRevX.6.031007}%
  \BibitemOpen
  \bibfield  {author} {\bibinfo {author} {\bibfnamefont {P.~J.~J.}\
  \bibnamefont {O'Malley}} \emph {et~al.},\ }\bibfield  {title} {\bibinfo
  {title} {Scalable quantum simulation of molecular energies},\ }\href
  {https://doi.org/10.1103/PhysRevX.6.031007} {\bibfield  {journal} {\bibinfo
  {journal} {Phys. Rev. X}\ }\textbf {\bibinfo {volume} {6}},\ \bibinfo {pages}
  {031007} (\bibinfo {year} {2016})}\BibitemShut {NoStop}%
\bibitem [{\citenamefont {Nam}\ \emph {et~al.}(2020)\citenamefont {Nam} \emph
  {et~al.}}]{ISI:000524530000001}%
  \BibitemOpen
  \bibfield  {author} {\bibinfo {author} {\bibfnamefont {Y.}~\bibnamefont
  {Nam}} \emph {et~al.},\ }\bibfield  {title} {\bibinfo {title} {Ground-state
  energy estimation of the water molecule on a trapped-ion quantum computer},\
  }\href {https://doi.org/10.1038/s41534-020-0259-3} {\bibfield  {journal}
  {\bibinfo  {journal} {npj Quantum Inform.}\ }\textbf {\bibinfo {volume}
  {6}},\ \bibinfo {pages} {33} (\bibinfo {year} {2020})}\BibitemShut {NoStop}%
\bibitem [{\citenamefont {Gao}\ \emph {et~al.}(2021)\citenamefont {Gao},
  \citenamefont {Nakamura}, \citenamefont {Gujarati}, \citenamefont {Jones},
  \citenamefont {Rice}, \citenamefont {Wood}, \citenamefont {Pistoia},
  \citenamefont {Garcia},\ and\ \citenamefont
  {Yamamoto}}]{doi:10.1021/acs.jpca.0c09530}%
  \BibitemOpen
  \bibfield  {author} {\bibinfo {author} {\bibfnamefont {Q.}~\bibnamefont
  {Gao}}, \bibinfo {author} {\bibfnamefont {H.}~\bibnamefont {Nakamura}},
  \bibinfo {author} {\bibfnamefont {T.~P.}\ \bibnamefont {Gujarati}}, \bibinfo
  {author} {\bibfnamefont {G.~O.}\ \bibnamefont {Jones}}, \bibinfo {author}
  {\bibfnamefont {J.~E.}\ \bibnamefont {Rice}}, \bibinfo {author}
  {\bibfnamefont {S.~P.}\ \bibnamefont {Wood}}, \bibinfo {author}
  {\bibfnamefont {M.}~\bibnamefont {Pistoia}}, \bibinfo {author} {\bibfnamefont
  {J.~M.}\ \bibnamefont {Garcia}},\ and\ \bibinfo {author} {\bibfnamefont
  {N.}~\bibnamefont {Yamamoto}},\ }\bibfield  {title} {\bibinfo {title}
  {Computational investigations of the lithium superoxide dimer rearrangement
  on noisy quantum devices},\ }\href {https://doi.org/10.1021/acs.jpca.0c09530}
  {\bibfield  {journal} {\bibinfo  {journal} {The Journal of Physical Chemistry
  A}\ }\textbf {\bibinfo {volume} {125}},\ \bibinfo {pages} {1827} (\bibinfo
  {year} {2021})},\ \bibinfo {note} {pMID: 33635672}\BibitemShut {NoStop}%
\bibitem [{\citenamefont {Bojowald}(2015)}]{Bojowald_2015}%
  \BibitemOpen
  \bibfield  {author} {\bibinfo {author} {\bibfnamefont {M.}~\bibnamefont
  {Bojowald}},\ }\bibfield  {title} {\bibinfo {title} {Quantum cosmology: a
  review},\ }\href {https://doi.org/10.1088/0034-4885/78/2/023901} {\bibfield
  {journal} {\bibinfo  {journal} {Reports on Progress in Physics}\ }\textbf
  {\bibinfo {volume} {78}},\ \bibinfo {pages} {023901} (\bibinfo {year}
  {2015})}\BibitemShut {NoStop}%
\bibitem [{\citenamefont {Liu}\ \emph {et~al.}(2021)\citenamefont {Liu},
  \citenamefont {Wu}, \citenamefont {Wan}, \citenamefont {Pan}, \citenamefont
  {Qin}, \citenamefont {Gao},\ and\ \citenamefont {Wen}}]{Liu_2021}%
  \BibitemOpen
  \bibfield  {author} {\bibinfo {author} {\bibfnamefont {H.-L.}\ \bibnamefont
  {Liu}}, \bibinfo {author} {\bibfnamefont {Y.-S.}\ \bibnamefont {Wu}},
  \bibinfo {author} {\bibfnamefont {L.-C.}\ \bibnamefont {Wan}}, \bibinfo
  {author} {\bibfnamefont {S.-J.}\ \bibnamefont {Pan}}, \bibinfo {author}
  {\bibfnamefont {S.-J.}\ \bibnamefont {Qin}}, \bibinfo {author} {\bibfnamefont
  {F.}~\bibnamefont {Gao}},\ and\ \bibinfo {author} {\bibfnamefont {Q.-Y.}\
  \bibnamefont {Wen}},\ }\bibfield  {title} {\bibinfo {title} {Variational
  quantum algorithm for the poisson equation},\ }\bibfield  {journal} {\bibinfo
   {journal} {Physical Review A}\ }\textbf {\bibinfo {volume} {104}},\ \href
  {https://doi.org/10.1103/physreva.104.022418} {10.1103/physreva.104.022418}
  (\bibinfo {year} {2021})\BibitemShut {NoStop}%
\bibitem [{\citenamefont {Sato}\ \emph {et~al.}(2021)\citenamefont {Sato},
  \citenamefont {Kondo}, \citenamefont {Koide}, \citenamefont {Takamatsu},\
  and\ \citenamefont {Imoto}}]{Sato_2021}%
  \BibitemOpen
  \bibfield  {author} {\bibinfo {author} {\bibfnamefont {Y.}~\bibnamefont
  {Sato}}, \bibinfo {author} {\bibfnamefont {R.}~\bibnamefont {Kondo}},
  \bibinfo {author} {\bibfnamefont {S.}~\bibnamefont {Koide}}, \bibinfo
  {author} {\bibfnamefont {H.}~\bibnamefont {Takamatsu}},\ and\ \bibinfo
  {author} {\bibfnamefont {N.}~\bibnamefont {Imoto}},\ }\bibfield  {title}
  {\bibinfo {title} {Variational quantum algorithm based on the minimum
  potential energy for solving the poisson equation},\ }\bibfield  {journal}
  {\bibinfo  {journal} {Physical Review A}\ }\textbf {\bibinfo {volume}
  {104}},\ \href {https://doi.org/10.1103/physreva.104.052409}
  {10.1103/physreva.104.052409} (\bibinfo {year} {2021})\BibitemShut {NoStop}%
\bibitem [{\citenamefont {Frazier}\ \emph {et~al.}(2019)\citenamefont
  {Frazier}, \citenamefont {Henderson},\ and\ \citenamefont
  {Waeber}}]{Waeber2016}%
  \BibitemOpen
  \bibfield  {author} {\bibinfo {author} {\bibfnamefont {P.~I.}\ \bibnamefont
  {Frazier}}, \bibinfo {author} {\bibfnamefont {S.~G.}\ \bibnamefont
  {Henderson}},\ and\ \bibinfo {author} {\bibfnamefont {R.}~\bibnamefont
  {Waeber}},\ }\bibfield  {title} {\bibinfo {title} {Probabilistic bisection
  converges almost as quickly as stochastic approximation},\ }\bibfield
  {journal} {\bibinfo  {journal} {Mathematics of Operations Research}\ }\textbf
  {\bibinfo {volume} {44}},\ \href
  {https://doi.org/https://doi.org/10.1287/moor.2018.0938}
  {https://doi.org/10.1287/moor.2018.0938} (\bibinfo {year} {2019})\BibitemShut
  {NoStop}%
\bibitem [{\citenamefont {Barends}\ \emph {et~al.}(2016)\citenamefont {Barends}
  \emph {et~al.}}]{Barends_2016}%
  \BibitemOpen
  \bibfield  {author} {\bibinfo {author} {\bibfnamefont {R.}~\bibnamefont
  {Barends}} \emph {et~al.},\ }\bibfield  {title} {\bibinfo {title} {Digitized
  adiabatic quantum computing with a superconducting circuit},\ }\href
  {https://doi.org/10.1038/nature17658} {\bibfield  {journal} {\bibinfo
  {journal} {Nature}\ }\textbf {\bibinfo {volume} {534}},\ \bibinfo {pages}
  {222} (\bibinfo {year} {2016})}\BibitemShut {NoStop}%
\end{thebibliography}%

\end{document}